\input{aipcheck}

\documentclass[
    ,final            % use final for the camera ready runs
  ]
  {aipproc}

\layoutstyle{6x9}

\begin{document}

\title{Seeking the progenitors of Type Ia Supernovae}


\author{F. Patat}{
  address={European Organization for Astronomical Research in the Southern 
	Hemisphere, K. Schwarzschild Str.2 85748 Garching b. M\"unchen, Germany}
}

\begin{abstract}
The nature of the progenitor system[s] of Type Ia Supernovae is still
unclear. In this contribution I review the projects that have been
undertaken to answer this question and the results they have led
to. The conclusion is that, as of today, we have reasonable guesses
but none of them has yet been proven by direct observations.
\end{abstract}

\maketitle

%%%%%%%%%%%%%%%%%%%%%%%%%%%%%%%%%%%%%%%%%%%%
%% MAINMATTER
%%%%%%%%%%%%%%%%%%%%%%%%%%%%%%%%%%%%%%%%%%%%

\section{The quest for the progenitor's nature}

{\it The fact that we do not know yet what are the progenitor system of
some of the most dramatic explosions in the universe has become a
major embarrassment and one of the key unresolved problems in stellar
evolution.} \citep{livio}.

\vspace{2mm}

Due to their enormous luminosities and their homogeneity, Type Ia
Supernovae (hereafter SN Ia) have been used in cosmology as reference
beacons, with the ambitious aim of tracing the evolution of the
universe \cite{riess,perlmutter}. Despite of the progresses made in
this field, the nature of the progenitor stars and the physics which
governs these powerful explosions are still uncertain
\cite{hillebrandt}.  In general, they are thought to originate from a
close binary system
\cite{whelan}, where a white dwarf accretes material from a companion
until it approaches the Chandrasekhar limit and finally undergoes a
thermonuclear explosion. This scenario is widely accepted, but the
nature of both the accreting and the donor star is not yet known, even
though favorite configurations do exist
\cite{branch,partha,tutukov}. But why is it so important to
investigate the nature of the progenitor system? Besides the
fundamental implications on the cosmological usage of SNe Ia, there
are actually several other reasons to bother \cite{livio}. First of
all, galaxy evolution depends on the radiation, kinetic energy and
nucleosynthesis yields of these powerful events. Secondly, the
knowledge of the initial conditions of the exploding system is crucial
for understanding the physics of the explosion itself. Finally,
identifying the progenitors and determining the SN rates will allow us
to put constraints on the theory of binary star evolution.

Having in mind why we want to do this, the next question is, as usual,
how. A discriminant between some of the proposed scenarios would be
the detection of circumstellar material (CSM). However,
notwithstanding the importance of the quest, all attempts of detecting
direct signatures of the material being transferred to the accreting
white dwarf in normal SNe~Ia were so far frustrated, and only upper
limits to the mass loss rate could be placed from optical
\cite{lundqvist03,lundqvist05,mattila}, radio \cite{panagia06} 
and UV/X-Ray emission \cite{immler}. Claims of possible ejecta-CSM
interaction have been made for a few normal objects, namely SN~1999ee
\cite{mazzali05}, SN2001el \cite{wang03}, SN~2003du \cite{gerardy} and
SN~2005cg \cite{quimby06}. In all those cases, the presence of CSM is
inferred by the detection of high velocity components in the SN
spectra. However, it must be noticed that these features can be
explained by a 3D structure of the explosion \cite{mazzali05} and,
therefore, circumstellar interaction is not necessarily a unique
interpretation \cite{quimby06}.

Two remarkable exceptions are represented by the peculiar SNe 2002ic
and SN 2005gj, which have shown extremely pronounced hydrogen emission
lines \cite{hamuy,aldering,prieto}, that have been interpreted as a
sign of strong ejecta-CSM interaction.  However, the classification of
these supernovae as SNe Ia has been questioned \cite{benetti06}, and
even if they were SN Ia, they must be rare and hence unlikely to
account for normal Type Ia explosions \cite{panagia06}. As a matter of
fact, the only genuine detection so far seemed to be represented by
the underluminous SN~2005ke \cite{patat05}, for which an unprecedented
X-ray emission at a 3.6$\sigma$-level accompanied by a large UV excess
was reported \cite{immler}. These facts have been interpreted as the
signature of a possible, weak interaction between the SN ejecta and
material lost by the donor and used to derive some of its properties
\cite{immler}. However, a thorough re-analysis of the data has not
confirmed these findings \cite{hughes}, thus reconciling the picture
with the lack of detection reported at radio wavelengths
\cite{soderberg}.

Another promising path that has been explored is the one of the
so-called entrained material. Very briefly, the impact of the SN
ejecta on the companion star is expected to strip its envelope. Part
of it becomes entrained in the ejected material and should be
observable at late phases in the form of narrow emission lines
\cite{wheel75,fryxell,taam,chugai86,livne,marietta,meng,pakmor}. 
So far, no trace of hydrogen (or helium) has been detected in the late
spectra of Type Ia SNe and this rules out systems with secondary stars
close enough to the exploding WD to be experiencing Roche-lobe
overflow at the time of explosion \cite{leonard}.

All the channels explored so far to detect CSM around Type Ia SN
progenitors are based on the fact that sooner or later the fast SN
ejecta will crash into the slow moving material lost by the system in
the pre-explosion phases in the form of stellar wind. This implicitly
requires two conditions to be fulfilled: i) there has to be
interaction and ii) the amount of CSM and its density must reach some
threshold values in order to produce a detectable
interaction. Therefore, methods based on ejecta-CSM interaction will
not be able to reveal this material if its amount is small, if it is
placed rather far from the explosion site or if nova-like evacuation
mechanisms are at work \cite{wood-vasey}.

So, the question is whether there exists any other method that does
not require interaction with the CSM. 

One of them has been proposed and applied very recently and, at
variance with those I have described so far, it targets at observing
the progenitor system before the explosion \cite{voss}. The idea
behind this new technique is that different progenitor systems should
have different properties when observed in the X-ray domain. More
precisely, while in a single-degenerate system X-ray emission is
supposed to be a by-product of the accretion process
\cite{vdheuvel}, this is most likely not the case for the alternative
scenario, where two WDs (henceforth also indicated as
double-degenerate) loose momentum through the emission of
gravitational waves and eventually merge, producing a thermonuclear
explosion \cite{iben,webbink}.  Clearly, in order to be able to apply
this method, one needs to have deep, pre-explosion X-ray images of the
explosion site, to be compared with analogous images taken after the
explosion has taken place.  In fact, the detection of an X-ray source
at a projected position coincident (to within the errors) with that of
the SN is not sufficient to conclude that the X-ray actually came from
the progenitor. It is only the disappearance of that source that would
strongly support this conclusion.

This idea has been applied for the first time to the nearby
SN~2007on. Based on pre-explosion Chandra archival data, Voss \&
Neelemans \cite{voss} have reported a 4 $\sigma$-level detection of an
X-ray source at a position which was consistent with that of the
SN. The implied luminosity is fully compatible with those of
super-soft sources \cite{cowley}, which have been identified as one of
the possible channels to Type Ia explosion \cite{vdheuvel}. In
general, the detection of a bright X-ray source spatially coincident
with SN~2007on has been interpreted in support the single-degenerate
accretion scenario, at least for this particular SN
\cite{voss}.

The site of SN~2007on was re-visited by Chandra about six weeks after
the SN reached maximum light and the astrometry of the field refined
\cite{roelofs}. After this re-examination two main facts have emerged:
i) there is a statistically significant offset between the
pre-explosion X-ray source and the SN position; ii) the X-ray source
is still there after the SN explosion, but its luminosity has dimmed
by about a factor four \cite{roelofs}. In the light of the fact that
no Type Ia SN has ever been detected in X-rays at any epoch
\cite{immler}, this indeed questions the conclusions originally
inferred for the nature of the source and its identification as the
progenitor of 2007on. Clearly, only further, deep Chandra observations
will allow us to reach a firmer conclusion \cite{roelofs}. This
technique has been applied to three other nearby Type Ia SN, for which
pre-explosion Chandra observations were available, but no X-ray source
was detected at the SN positions \cite{voss}.  Even though its
application has the availability of archival, X-ray data as a
mandatory pre-requisite, it is certainly a promising tool to
investigate the nature of Type Ia progenitors.

\section{Not through Hydrogen: yet another method to detect CSM in Ia's}

\begin{figure}
\resizebox{.9\textwidth}{!}
{\includegraphics{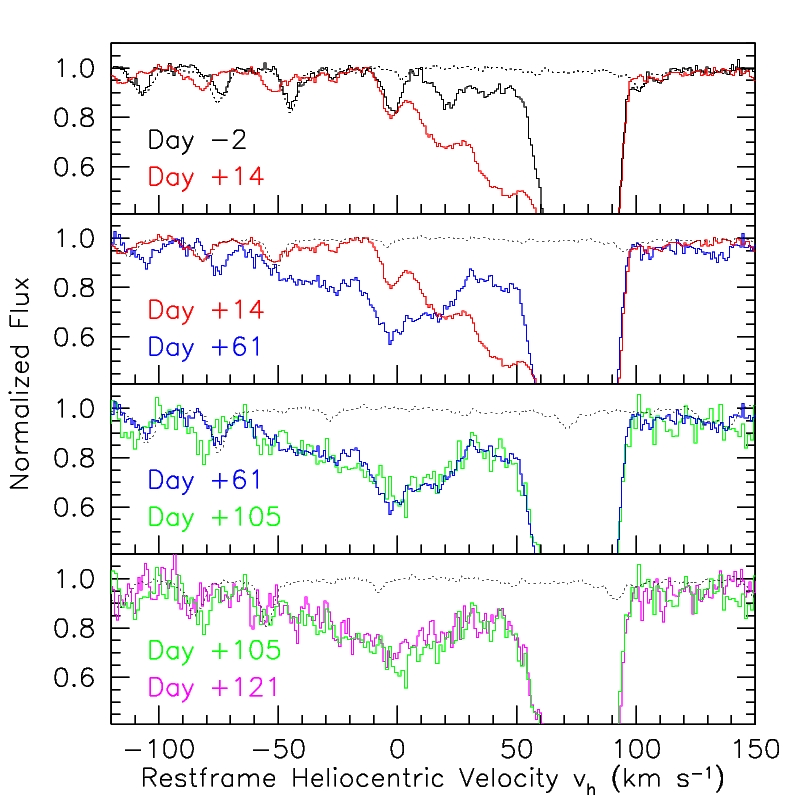}}
\caption{Time evolution of the sodium D$_2$ component region as a function
of elapsed time since B-band maximum light. In each panel the dotted
curve traces the atmospheric absorption spectrum.}
\label{fig:naevol}
\end{figure}

While investigating the effects of a close-by, dusty environment on
the light curves and spectra of Type Ia SNe \cite{patat05b,patat06},
we saw another possibility of revealing possible CSM without the need
of having matter interaction.  In fact, since in SNe Ia the UV flux
bluewards of 350 nm undergoes severe line blocking by heavy elements
like Fe, Co, Ti and Cr \cite{pauldrach,mazzali00}, they are able of
ionizing possible CSM only within a relatively small radius. Once the
UV flux has significantly decreased in the post-maximum phase, if the
material has a sufficiently high density it recombines, producing
time-variable absorption features. Of course, if the material where
these features arise is reached by the fast moving ejecta, it will be
shocked and ionised, causing the disappearance of such
absorptions. Among all possible inter/circumstellar absorption lines,
the ubiquitous Na~I D lines are the best candidates for this kind of
study. In fact, besides falling in an almost telluric absorption-free
spectral region, they are produced by a very strong transition, and
hence detectable for rather small gas column densities. In addition,
the ionization potential of Na I is low (5.1 eV), and this ensures
that even a weak UV field is able to have a measurable effect on its
ionisation, without the need for direct interaction.

With this idea in mind, the experimental path was rather clearly
traced: obtain multi-epoch, high-resolution spectroscopy of the next
bright SN Ia and look for absorption-line variability.  The first
chance to test our idea came when SN~2006X was discovered in the Virgo
Cluster spiral galaxy M100 on February 4, 2006. A few days later, the
object was classified as a normal Type Ia event occurring 1--2 weeks
before maximum light and suffering substantial extinction
\cite{quimby} . Prompt Very Large Array (VLA) observations have shown
no radio source at the SN position, establishing one of the deepest
and earliest limits for radio emission from a Type Ia, and implying a
mass-loss rate of less than a few 10$^{-8}$ M$_\odot$ yr$^{-1}$ (for a
wind velocity of 10 km s$^{-1}$) \cite{stockdale}. The SN was not
visible in the 0.2-10 keV X-rays band down to the SWIFT satellite
detection limit \cite{immler}. All of this made of SN2006X a perfect
candidate to verify our idea.  The observations started on February
18th and were carried out with the Ultraviolet and Visual Echelle
Spectrograph (UVES) mounted at the Very Large Telescope on four
different epochs, which correspond to days $-$2, +14, +61 and +121
with respect to B-band maximum light.  Additionally, a fifth epoch
(day +105) was covered with the High Resolution Echelle Spectrometer
mounted at the 10m Keck telescope. The data show a wealth of
interstellar features
\cite{cox}, but the most remarkable finding is the clear evolution
seen in the profile of the Na I D lines
\cite{patat}. In fact, besides a strongly saturated and constant
component, arising in the host galaxy disk, a number of features
spanning a velocity range of about 100 km s$^{-1}$ appear to vary
significantly with time (see Figure~\ref{fig:naevol}). SN2006X is
projected onto the receding side of the galaxy, and the component of
the rotation velocity along the line of sight at the apparent SN
location is about +75 km s$^{-1}$, which coincides with the strongly
saturated NaI D component, the saturated Ca II H\&K lines, and a
weakly saturated CN vibrational band.  This and the lack of time
evolution proves that the deep absorption arises within the disk of
M100 in an interstellar molecular cloud (or system of clouds) that is
responsible for the bulk of reddening suffered by SN2006X.  In
contrast, the relatively blue-shifted structures of the Na I D lines
show a rather complex evolution. Without going into the details (the
reader is referred to \cite{patat}), these findings were interpreted
as a solid evidence of CSM expanding at velocities that span a range
of about 100 km s$^{-1}$, placed between 10$^{16}$ and 10$^{17}$ cm
from the explosion site and ejected from the progenitor system in the
recent past. This almost certainly rules out a double-degenerate
scenario for SN2006X, where the supernova would have been triggered by
the merger of two CO white dwarfs. In this case, no significant mass
loss would be expected in the phase immediately preceding the
supernova and, thus, a single-degenerate system is the favored one for
SN 2006X\footnote{However, I think it is fair to report here a private
communication I received from A.V. Tutukov (May 2008). He maintains
that in a binary system formed by two degenerate stars, the common
envelope can survive for about 10$^4$ years. This time is sufficient
for the two WDs to merge via gravitational waves radiation if the
separation between the two components is within a suitable
range. Based on his calculations, he concludes that about one third of
all Ia could form in double degenerate systems still embedded in their
common envelopes and, therefore, produce absorption features similar
to those seen in SN~2006X at the time of explosion. For this reason,
he reckons, these findings are not necessarily excluding a double
degenerate system for this particular event.}.

The observed structure of the circumstellar material could be due to
variability in the wind from the companion red giant, since
considerable variability of red giant mass loss is generally expected
\cite{willson}.  A potentially more interesting interpretation of
these distinct features is that they arise in the remnant shells (or
shell fragments) of successive novae, which can create over-density
regions in the slow moving material released by the companion, also
evacuating significant volumes around the progenitor star
\cite{judge,hachisu,wood-vasey}. The calculations have difficulty in
matching the velocities in our observations if the nova shells are
decelerated in a spherically symmetric wind, in the sense that the
measured velocities are too small. However, if the wind is
concentrated towards the orbital plane this discrepancy could be
removed, since the nova shell would be more strongly decelerated in
the equatorial plane; in that case we would be observing the supernova
close to the orbital plane. Not only might this be expected a priori,
but observations of the 2006 outburst of RS Ophiuchi show that the
nova ejecta are bipolar and that there is an equatorial density
enhancement which strongly restrains the expansion of the nova shell
\cite{obrien}, providing some support for such a scenario.

Interestingly, short-lived heavy element absorption systems have been
reported for a number of novae near maximum light and interpreted as
the signature of circumbinary material \cite{williams}. It has also
been suggested that very early multi-epoch, high resolution
spectroscopy of Type Ia SNe might be used to disentangle between
different progenitor scenarios through the study of those systems
\cite{williams}. Clearly, in the case of a nova, the spatial scales
that can be probed by the observations are much smaller than in the
case of a SN, where the fast expanding ejecta sweep some 20 AU during
the first day after the explosion. This, together with the fact that a
SN is typically discovered several days after the explosion, makes it
very difficult to study the progenitor system down to circumbinary
scales. Even in the hypothesis this material has not yet been reached
by the ejecta, being so close to the explosion it is most likely
completely ionised by the SN radiation field.

\section{Problems and Perspectives}

One problem with the scenario proposed by Patat et al. \cite{patat} to
explain the behavior seen in SN~2006X is that even though the UV field
produced by a Type Ia explosion is relatively weak, the ionisation
rate is still sufficiently high to keep sodium fully ionised in a
spherically symmetric wind produced by a red giant for more than a
month at distances smaller than 10$^{17}$ cm. This implies that the
time-evolving Na~I D lines detected in 2006X cannot arise in the wind
of a red giant star \cite{chugai}. For this reason, N. Chugai has
proposed that these features have nothing to do with the circumstellar
environment, arise in clouds placed at distances larger than 10$^{17}$
cm, and their variations are due to a purely geometrical effect
\cite{chugai}. This possibility was discussed and excluded by Patat et
al. \cite{patat}, based on the fact that while the Na~I D features
vary, Ca~II H\&K do not. It must be noticed that, in order for the
geometrical effect to explain the observed behavior, a rather ad-hoc
cloud geometry and chemical composition (and/or physical conditions)
are required \cite{chugai}. One prediction that follows from the
scenario proposed by Chugai is that these time-variant features should
have a higher probability of being detected in highly reddened events
\cite{chugai}.  In this respect, it is interesting to note that a
study conducted on a sample of low resolution spectra of 31 Type Ia
SNe has shown that only two SNe displayed Na~I D variability, namely
SNe 1999cl and 2006X, the two most highly reddened objects of the
sample \cite{blondin}. However, another highly reddened SN included in the
sample (SN~2003cg) did not show any trace of variability.

Even though the results obtained with multi-epoch, high-resolution
observations of SN~2006X have already triggered a couple of similar
studies \cite{patatcx,simon}, the sample is simply too small to allow
any firm conclusion. For this purpose we have started a program at the
VLT with the aim of obtaining data for a fair number of nearby events.
During the first year of activity three SNe have been observed, i.e.
2008ec, 2008fp and 2008hv. The data will be presented soon, but here I
can anticipate that none of them has shown any signs of variability,
even if the first two objects suffered a substantial extinction within
the host galaxy, as witnessed by saturated Na~I D lines.  So far, only
another case, even though not so extreme as SN~2006X, has been found
\cite{simon09}. In fact, multi-epoch, high resolution spectroscopy of
SN~2007le has clearly shown variability in the Na~I D profiles, while
the Ca~II H\&K profiles did not show any
evolution\footnote{Unfortunately, nothing can be said about the
behavior of Ca~II H\&K in SN~1999cl, the only other known case of
variability \cite{blondin}.}. This dicothomy in the behavior of Na~I
vs. Ca~II lines is probably the key fact in the understanding of what
the physical reason for the observed variability is. If more cases are
found, the chances this is due to a fortuitous series of coincidences
will become negligibly small. What we have seen in 2006X is far from
being completely understood and we are certainly left with more
questions than answers.

After reviewing all the methods that have been proposed and applied so
far to directly\footnote{Note that there are other, less direct paths
to approach this problem, based for instance on the chemical yields,
delay times and rates of Type Ia explosions, which I have not
considered here. See for instance \cite{greggio} and references
therein.}  unveil the nature of Type Ia SN progenitors, it is clear
that none of them has given a satisfactory picture and the systems
which give rise to one of the most catastrophic events in the universe
remain concealed. The only thing we know is that they must be
surrounded by tiny amounts of material, otherwise we would have
detected them through the interaction with the SN ejecta. Whether this
gas is in the form of dense clumps, an equatorial torus, nested and
fragmented shells or bipolar structures remains to be clarified.

I believe that by now we are all convinced that there is more than one
channel leading to the same explosive theme, on top of which nature
adds some variations, as the non perfect homogeneity of SNe Ia seems
to tell us. Rather than the end of an old story, I consider the most
recent findings in this field as the beginning of a new one.

%%%%%%%%%%%%%%%%%%%%%%%%%%%%%%%%%%%%%%%%%%%%%%%%
%% BACKMATTER
%%%%%%%%%%%%%%%%%%%%%%%%%%%%%%%%%%%%%%%%%%%%%%%%

\begin{theacknowledgments}
I wish to thank the organizing committee for inviting me to give this
review. I also like to thank all my collaborators, for supporting and
encouraging me throughout this project.
\end{theacknowledgments}



\bibliographystyle{aipprocl} % if natbib is missing

\begin{thebibliography}{9}
\bibitem{aldering} G. Aldering et al., 2006, ApJ, 650, 510
\bibitem{benetti06} S. Benetti et al., 2006, ApJL, 653, L129
\bibitem{blondin} S. Blondin et al., 2008, ApJ in press (arXiv:0811.0002)
%\bibitem{bode} Bode, M.F. et al., 2007, ApJL, in press (arXiv:0706.2745)
\bibitem{branch94} D. Branch \& S. van den Bergh, 1993, AJ, 105, 2231
\bibitem{branch} D. Branch, M. Livio, \& L.R. Yungelson,
        F.R. Boffi \& E. Baron, 1995, PASP,  107, 1019
\bibitem{chugai86} N.N. Chugai, 1986,SvA, 30, 563--566
\bibitem{chugai} N.N. Chugai, 2007, Astronomy Letters, 34, 389--396
\bibitem{cowley} A.P. Cowley et al., 1998, ApJ, 504, 854--865
\bibitem{cox} N. Cox \& F. Patat, 2008, A\&A,  485, L9--12
\bibitem{fryxell} B.A. Fryxell \& W.D. Arnett, 1981, ApJ, 243, 994, 1002
\bibitem{gerardy}C.L. Gerardy, et al., 2004, ApJ, 607, 391
\bibitem{greggio} L. Greggio, A. Renzini \& E. Daddi, 2008, MNRAS, 388, 
	829--837
\bibitem{hamuy} M. Hamuy et al., 2003, Nature, 424, 651
\bibitem{hillebrandt}] W. Hillebrandt \& J.C. Niemeyer, 2000, ARAA, 38, 191
%\bibitem{han} Han Z. \& Podsiadlowski Ph., 2004, MNRAS, 350, 1301
\bibitem{hughes} J.P. Hughes, N.N. Chugai, R. Chevalier, P. Lundqvist 
	\& E. Schlegel, 2007, ApJ, 670, 1260
\bibitem{iben} I. Iben \& A.V. Tutukov, 1984, ApJS, 84, 335--372
\bibitem{immler} S.I. Immler et al., 2006, ApJ, 648, L119
\bibitem{judge} P.G. Judge \&  R.E. Stencel, 1991, ApJ, 371, 357
\bibitem{hachisu} I. Hachisu \& M. Kato, 2001, ApJ, 558, 323
\bibitem{leonard} D.C. Leonard, 2007, ApJ, 670, 1275--1282
\bibitem{livne} E. Livne, Y. Tuchman \& J.C. Wheeler, 1992, ApJ, 399, 665--671
\bibitem{livio} M. Livio, 2001, in Supernovae and gamma-ray bursts: the 
        greatest 
	explosions since the Big Bang. Proceedings of the STScI Symposium, 
	edited by M. Livio, N. Panagia and K. Sahu. STScI symposium series,
	Vol. 13. Cambridge, UK: Cambridge University Press, p. 334--355
\bibitem{lundqvist03} P. Lundqvist et al., 2003, in
        {\it From twilight to highlight: the physics of Supernovae},
        ed. W. Hillebrandt \& B. Leibundgut (Berlin: Springer), 309
\bibitem{lundqvist05} P. Lundqvist et al., 2005, in {\it Cosmic
        Explosions}, ed. J. M. Marcaide, \& K. W. Weiler, CD-ROM version,
        IAU Coll., 192, 81
\bibitem{marietta}E. Marietta, A. Burrows \& B. Fryxell, 2000, ApJ,
	128, 615--650
\bibitem{mattila} S. Mattila et al., 2005,  A\&A, 443, 649
\bibitem{mazzali00} P.A. Mazzali et al., 2000, A\&A, 363, 705
\bibitem{mazzali05} P.A. Mazzali et al., 2005, MNRAS, 357, 200
\bibitem{meng} X. Meng, X. Chen \& Z. Han, 2007, PASJ, 59, 835--840
\bibitem{obrien} T.J. O'Brien et al., 2006, Nature, 442, 279
\bibitem{pakmor}R. Pakmor, F.K. R\"opke, A. Weiss \& W. Hillebrandt, 2008,
	489, 943--951
\bibitem{panagia06} Panagia, N., et al., 2006, ApJ, 469, 396
\bibitem{partha} M. Parthasarathy, D. Branch, D.J. Jeffery \&
        E. Baron, 2007, New Astronomy Reviews, 51, 524
\bibitem{patat05} F. Patat, D. Baade, L. Wang, S. Taubenberger \&
        J.C. Wheeler, 2005, IAU Circ. n.~8631
\bibitem{patat05b} F. Patat, 2005, MNRAS, 357, 1161--1177
\bibitem{patat06} F. Patat, S. Benetti, E. cappellaro \& M. Turatto, 2006,
	MNRAS, 369, 1949--1960
\bibitem{patat} F. Patat et al., 2007, Science, 317, 924
\bibitem{patatcx} F. Patat et al., 2007, A\&A, 474, 931--936

\bibitem{pauldrach} W.A. Pauldrach et al., 1996, A\&A, 312, 525
\bibitem{perlmutter} S. Perlmutter et al., 1999, ApJ, 517, 565
\bibitem{prieto} J.L. Prieto et al., 2008, ApJ, submitted
        (arXiv:0706.4088)
\bibitem{quimby06} R. Quimby, et al., 2006, ApJ, 636, 400
\bibitem{quimby} R. Quimby, P. Brown \& C. Gerardy,2006, CBET 421.
\bibitem{rappaport} S. Rappaport, R. Di Stefano \& J.D. Smith,
        1994, ApJ, 426, 692
\bibitem{riess} A.G. Riess et al., 1998, AJ, 116, 1009
\bibitem{roelofs} G. Roelofs, C. Bassa, R. Voss \& G. Nelemans, 2008, 
	MNRAS, 391, 290
\bibitem{simon} J.D. Simon et al., 2007, ApJ, 671, L25--28
\bibitem{simon09} J.D. Simon et al., 2009, in preparation
\bibitem{soderberg}A.M. Soderberg, 2006, Astron. Tel., 722, 1
\bibitem{stockdale}C.J. Stockdale et al., 2006, CBET 396
\bibitem{taam} R.E. Taam \& B.A. Fryxell, 1984, ApJ, 279, 166--176
\bibitem{turatto} Turatto, M., Benetti, S. \& Cappellaro, E., 2003, in
        {\it From twilight to highlight: the physics of Supernovae},
        ed. W. Hillebrandt \& B. Leibundgut (Berlin: Springer), 200
\bibitem{tutukov} A.V. Tutukov \& A.V. Fedorova, 2007,
        Astronomy Reports, 51, 291
\bibitem{vdheuvel} E.P.J. van den Heuvel, D. Bhattacharya,
        K. Nomoto \& S. Rappaport, 1992, A\&A, 262, 97--105
\bibitem{voss} R. Voss \& G. Nelemans, 2008, Nature, 451, 802--804
\bibitem{wang03} L. Wang, et al., 2003, 591, 1110
\bibitem{webbink} R. Webbink, 1984, ApJ, 277, 355--360 
\bibitem{wheel75} J.C. Wheeler, M. Lecar \& C.F. McKee, 1975, ApJ, 200,
	145--157
\bibitem{whelan} J. Whelan \& I. Iben, 1973, ApJ, 186, 1007
\bibitem{williams} R. Williams, E. Mason, M. Della Valle \& A. Ederoclite,
	2008, ApJ, 685, 451--462
\bibitem{willson} L.A. Willson, L.A., 2000, ARAA, 38, 573
\bibitem{wood-vasey} W.M. Wood-Vasey \& J.L. Sokoloski, 2006,
        ApJ, 645, L53
%\bibitem{yu} Yu, C., Modjaz, M. \& Li, W.D., 2000, IAU Circ. 7458
\end{thebibliography}

\end{document}